\documentclass{llncs}

\usepackage{booktabs} 
\usepackage{hyperref}
\usepackage{multirow}
\usepackage{xcolor}
\usepackage{framed}
\usepackage{tablefootnote}
\usepackage{array}
\usepackage[utf8]{inputenc}
\usepackage{hyperref}
\usepackage{array,graphicx}
\usepackage{pxfonts}
\usepackage{tabularx}

\newcommand*\rot{\rotatebox{90}}

\newcommand{\removeadma}[1]{#1}

\begin{document}
\title{Doctoral Advisor or Medical Condition:\\ Towards Entity-specific Rankings\\ of Knowledge Base Properties\\ {[Extended Version]}\thanks{Conference version published at ADMA 2017}}
\author{Simon Razniewski$^{1}$, Vevake Baralaman$^2$, Werner Nutt$^1$}
\institute{$^1$Free University of Bozen-Bolzan \ \ \ \  $^2$University of Trento\\
\email{\{razniewski,nutt\}@inf.unibz.it, vvek.9291@gmail.com}}


\maketitle

\begin{abstract}
In knowledge bases such as Wikidata, it is possible to assert a large set of properties for entities, ranging from generic ones such as name and place of birth to highly profession-specific or background-specific ones such as doctoral advisor or medical condition. Determining a preference or ranking in this large set is a challenge in tasks such as prioritisation of edits or natural-language generation.  Most previous approaches to ranking knowledge base properties are purely data-driven, that is, as we show, mistake frequency for interestingness.

\removeadma{}
In this work, we have developed a human-annotated dataset of 350 preference judgments among pairs of knowledge base properties for fixed entities. From this set, we isolate a subset of pairs for which humans show a high level of agreement (87.5\% on average).  We show, however, that baseline and state-of-the-art techniques achieve only 61.3\% precision in predicting human preferences for this subset.
\removeadma{

We then analyze what contributes to one property being rated as more important than another one, and identify that at least three factors play a role, namely (i) general frequency, (ii) applicability to similar entities and (iii) semantic similarity between property and entity. We experimentally analyze the contribution of each factor and show that a combination of techniques addressing all the three factors achieves 74\% precision on the task. }
The dataset is available at\\ { \url{www.kaggle.com/srazniewski/wikidatapropertyranking}}.
\end{abstract}

%
%




\section{Introduction}

General-purpose knowledge bases such as Wikidata~\cite{wikidata}, YAGO~\cite{yago} or DBpedia~\cite{dbpedia} are becoming increasingly popular, and are used for a variety of tasks such as structured search, entity recognition or question answering. These knowledge bases can store a large number of entity types, and for each entity type a large number of properties. For instance, for the class of \emph{human} alone, more than 100 properties are used in Wikidata at least 1000 times, among which are the following:

\medskip
\hspace{1.4cm}\begin{tabular}{ll}
\label{tbl:intro}
\small 1: sex or gender \hspace{1.8cm} & \small 70: doctoral advisor\\
\small 2: occupation & \small 71: pseudonym\\
\small 3: date of birth & \small 72: medical condition\\
\small ... & \small ...\\
\small 39: height & \small 78: convicted of\\
\small 40: instrument & \small 79: singles record\\
\small ... & \small ...\\
\end{tabular}
\medskip

An issue with these properties is that it is not known how interesting they are for specific entities. For instance, while 90\% of the data in Wikidata is created by bots~\cite{wikidata}, it is not clear whether the entered data captures actually what is of interest to humans.
As a consequence, the large number of properties and their unclear interestingness severely hinder the usability of knowledge bases, and make many data analytics tasks difficult. 

A way to better structure these properties would be rankings by interestingness. Such rankings would be useful for at least three tasks:

\begin{enumerate}
\item \emph{Recommendations to authors:} Rankings by interestingness could help human authors in focusing their work~\cite{DBLP:journals/dbsk/AbedjanN13,ibminer,deswebe-2014-hpi}. For Wikidata, there exists a tool called \emph{Wikidata property suggester}\footnote{https://github.com/Wikidata-lib/PropertySuggester} \removeadma{(see Fig.~\ref{fig:prop-sugg})} for that purpose. However, the current instance is association-rule-based, which, as we show below, does not well approximate the human perception of interestingness.
\item \emph{Automatically generating descriptions:} One of the major motivations for the Wikidata project is to automatically generate article stubs, which is especially relevant for low-resource languages. As a 2015 report of the Wikimedia Foundation found, "[the] lack of any clear identifier for importance or primacy in Wikidata items"\footnote{\url{https://meta.wikimedia.org/wiki/Research:Wikidata_gap_analysis\#Conclusion}} is one of the primary obstacles to this goal.
\item The lack of ordering also makes comparing the relative completeness of entities~\cite{vision}, as for instance attempted by the Recoin tool~\cite{recoin}, difficult.
\end{enumerate}

\removeadma{
\begin{figure}
    \centering
    \includegraphics[width=0.6\columnwidth]{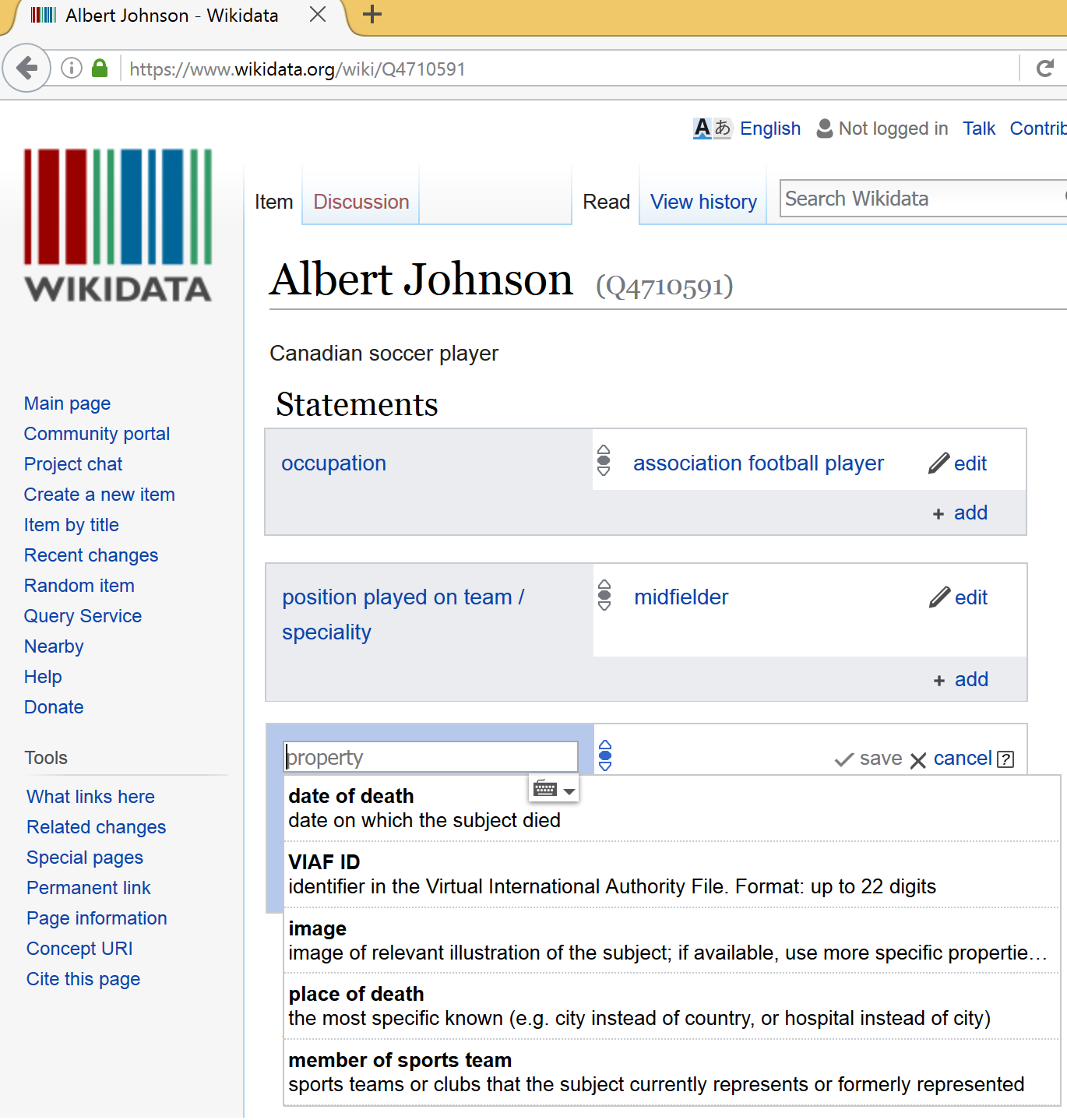}
    \caption{The current Wikidata property suggester.}
    \label{fig:prop-sugg}
\end{figure}
}

Predicting the interestingness of properties is difficult for at least three reasons: (i) Looking at the knowledge base alone is not sufficient: just because a property is very frequent in a knowledge base, one cannot conclude that the property is also very important.  For instance, there are about 27k people with a blood type, but only 2k people with a hair color in Wikidata, but nevertheless, the latter is generally more interesting than the former. (ii) The interestingness of properties is very dependent on the person.  For a politician, for instance, the political party is generally much more important than music instruments played, while for musicians, it is usually the other way around. (iii) There is a lack of datasets for this task, as most previous work used ablation studies, i.e., assessed performance on randomly removed portions of the data.

Previous work on property recommendation has mostly focused on data-driven approaches~\cite{innsbruck:property:ranking,DBLP:journals/dbsk/AbedjanN13,deswebe-2014-hpi,gassler2014guided}, which does not approximate human judgment too well. For instance, the Wikidata Property Suggester recommends to add \emph{date of death} and \emph{place of death} as most important missing properties to nearly all persons still alive. Similarly, it appears that frequent properties such as \emph{gender} and \emph{nationality} are overrated.
Closest to ours, work by Atzori and Dessi~\cite{dessi:atzori:ranking:properties} has investigated how to predict what human annotators actually find important, ignoring however the characteristics of individual entities, and using listwise learning-to-rank approaches that are not scalable.

\paragraph{Contribution}
Our technical contributions are:
(i) We introduce the problem of property ranking and discuss its significance in Section~\ref{chall}.
(ii) We develop a human-annotated gold-standard dataset containing 350 sets of an entity and two properties, each annotated with 10 preference judgments in Section~\ref{data}.
(iii) We evaluate baseline approaches and the state-of-the-art against our dataset, showing that these only achieve 61.3\% precision on records where humans have 87.5\% agreement (Section~\ref{sec:baselines}).
(iv) We develop techniques based on regression, LSI, LDA and ensembles that are able to achieve 74\% precision in Section~\ref{sec:advanced}.

\section{Background}


\paragraph{Learning to rank (L2R)} Learning to rank is a classic machine learning problem, where one aims to learn how to optimally rank given items. There are three main approaches to L2R, the so-called pointwise, pairwise, and listwise ranking~\cite{learningtorank,l2r2}. The pointwise approach, which is usually the easiest to implement, is based on the idea that each item has a score, which can be learned. Items can then be ranked by their score. Issues with the pointwise approach are mainly that the individual scores can be hard to interpret, 
and are not stable wrt.\ framing.

The pointwise and listwise approach aim to overcome this limitation of the pointwise approach by learning from ranked pairs and ranked lists, respectively. They can lead to more stable and better rankings, however, potentially require more effort during training data creation.

\removeadma{
\begin{figure}
\centering
    \includegraphics[width=0.7\columnwidth]{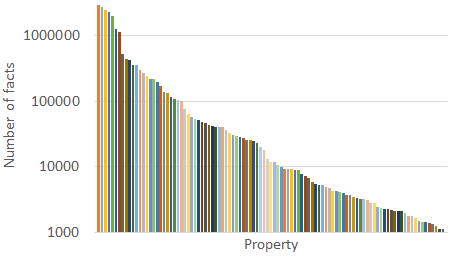}
    \caption{Frequency distribution of Wikidata properties used on humans at least 1000 times.}
    \label{fig:prop-distr}
\end{figure}
}

\paragraph{Wikidata} Wikidata is a crowd-sourced knowledge base that maintains information about entities of human knowledge, called \emph{items} in Wikidata parlance, which can be topics of Wikipedia articles (people, cities, movies, etc.) or anything else deemed of interest.
Information about an entity is structured as a collection of \emph{statements}, 
which are pairs consisting of a key, called, \emph{property}, and a value, 
which can be an atomic data value, an item, a property, or some possibly complex structure.
\removeadma{An entity can have an arbitrary number of statements with the same property.
Statements themselves can be further annotated with properties and values.

As properties are the main instrument to represent and structure information, 
introducing new properties into Wikidata is complicated,
while adding entities is easy. 
Any contributor is free to insert new entities
or add new statements to existing ones,
while new properties are only created after a lengthy discussion process in the Wikidata community.}
Currently, Wikidata contains roughly 25 mio.\ entities, and 2719 properties, whose usage follows roughly an exponential distribution.
%
This has two implications: First, it confirms the perception that many properties are quite specific, and apply only to few people. Second, it indicates that solely frequency-based rankings of properties tend to become imprecise in the long tail.

\paragraph{Ranking of Knowledge Base Properties}
There have been various works on knowledge base property ranking.
Most prominently, editors of Wikidata items are supported by the Property Suggester facility,
which, given an entity, produces a list of typically~3 to~10 properties 
for which no statement exists as yet.
It implements an approach by Abedjan and Naumann \cite{DBLP:journals/dbsk/AbedjanN13} that leverages techniques from association rule mining\removeadma{ \cite{Agrawal:Et:Al-Assoc:Rules-SIGMOD}}, and ranks rules according to squared confidence.
\removeadma{To check whether a new property $p'$ should be suggested for entity $e$,
confidence values for rules $p\to p'$, where $p$ is already a property of $e$,
are computed as usual \cite{Agrawal:Et:Al-Assoc:Rules-SIGMOD}.
Then $p'$ is suggested for $e$ if the sum of squared confidence values
exceeds a given threshold.}
On the one hand, this allows a few strong rules to outweigh many weaker ones,
while on the other hand inapplicable properties may be suggested 
if they are highly correlated with some existing properties.

Atzori and Dessi \cite{dessi:atzori:ranking:properties} addressed the problem to rank the existing properties of individual items in a knowledge base, like Wikidata or DBpedia. 
They identified three parameters for automating such rankings:
(i) algorithms for machine learning to rank;
(ii) possible features of properties; and 
(iii) methods to create training sets.
In a study, they chose eight algorithms, nine features, and six training
sets and combined them in all possible ways.
To test these combinations, they let a group of students pointwise rank the properties of 50 random Wikipedia entities. 
It turned out that the best combinations beat other state-of-the art techniques  by improvements of precision and recall of 5 to 10\%.
Like the authors of the present paper, Atzori and Dessi aim at creating 
automated methods that approximate human judgment about property ranks.
However, while they want to rank the properties that are already present for an item, we want to find out which properties of an item humans would find important or interesting, regardless of whether or not they are mentioned. Also, we aim to include information specific to the entity, not just use information on the level of the whole class.

\removeadma{
Other work on property ranking has attempted to reverse-en\-gi\-neer the structured data as shown by the Google search engine~\cite{reverse-engineer-properties}.
}

\paragraph{Fact Ranking}
Ranking knowledge base facts by importance, interestingness or unexpectedness is a very related topic~\cite{deswebe-2014-hpi,unusual-suspects,prakash2015did,gamon-text}.
Recent work by Bast et al.~\cite{bast:relevance:scores:triples}, for instance, investigates how to predict the relevance of attribute values on a scale from 0 to 1 for the multi-valued attributes profession and nationality. 
In their work, they show that methods that are based on the Wikipedia articles of persons and use a generative model can achieve reasonable accuracies on this task, i.e., less than 28\% numerical error in 80\% of cases. The problem was subsequently posed as challenge at the WSDM 2017 conference~\cite{wsdm-cup}.
\removeadma{}
However, having a ranking for facts does not help in the three applications above, as for the recommendation as to what to add, and the completeness comparison task, the properties that do not yet have facts are the ones that should be ranked, while for the natural language generation task, it is generally desired that values of multivalued properties appear together, thus, a ranking of individual facts will not do.

\removeadma{
\paragraph{Other}
Related is also work on schema learning. Schema learning aims to find out which properties are typical for a class or an entity in a given knowledge base~\cite{schema-induction,schema-induction-2-probabilistic,lee2013attribute}. However, the notion of typicality is usually based on statistics and in particular frequency and co-occurrence, which, as we show in Section~\ref{sec:baselines}, are not very good proxies for interestingness.
}

\section{Problem Definition and Challenges}
\label{chall}

We define our problem as follows:
\medskip

\noindent
\emph{Problem:} Given an entity and a set of knowledge base properties, rank the properties according to their interestingness for the entity.

\medskip

The notion of \emph{interestingness} is hereby left somewhat vague, which however is intentional as our ranking is not intended to serve only one specific purpose\removeadma{ (see also discussion in Section~\ref{sec:discussion})}.
We identify the following technical challenges for this task:
\begin{enumerate}
\item \emph{Lack of datasets}: Previous approaches~\cite{innsbruck:property:ranking,deswebe-2014-hpi,DBLP:journals/dbsk/AbedjanN13,gassler2014guided} rely on ablation studies, i.e., they randomly remove a portion of knowledge base facts, then evaluate how well they can predict removed facts or properties. This however does not say anything about their ability to rank properties by interestingness. The only dataset available is that of~\cite{dessi:atzori:ranking:properties}, which contains however only pointwise annotations for properties of 50 entities, and is of unknown quality.
\item \emph{Inapplicability of pointwise and listwise annotations:} While pointwise annotations are easy to solicit, the resulting scores can only be  interpreted within the context in which they were generated~\cite{pointwise-not-good,pointwise-not-good-2}. 
Using the annotation scheme from~\cite{bast:relevance:scores:triples}, for instance, we found that the 0-1 interestingness score of properties of medium interestingness was 0.43 when preceded by 7 questions about more interesting properties, but 0.62 if preceded by 7 less interesting ones\removeadma{ (more details in Appendix~\ref{appendix-frameproblem})}.
As the set of properties in Wikidata is constantly growing,\footnote{For instance, as of March 21st, 2016, there were 2202 properties, while as of February 7, 2017, there are 2719 according to \url{https://tools.wmflabs.org/hay/propbrowse/}} this would mean that pointwise scores collected now could not be interpreted well at later development stages of Wikidata.
%
%
Likewise, listwise approaches are not applicable, as it is not possible to elicit meaningful rankings for large sets of items.
\item \emph{No supervised learning possible:} 
Given that there are more than 3 million \emph{humans} in Wikidata, and that we rely on pairwise annotations, it is clear that it is impossible to generate enough training data for supervised learning.
\end{enumerate}
We next address Issue (1), the generation of a suitable dataset.

\section{Dataset Preparation}
\label{data}

\begin{figure*}[t]
    \centering
    \includegraphics[width=1\textwidth]{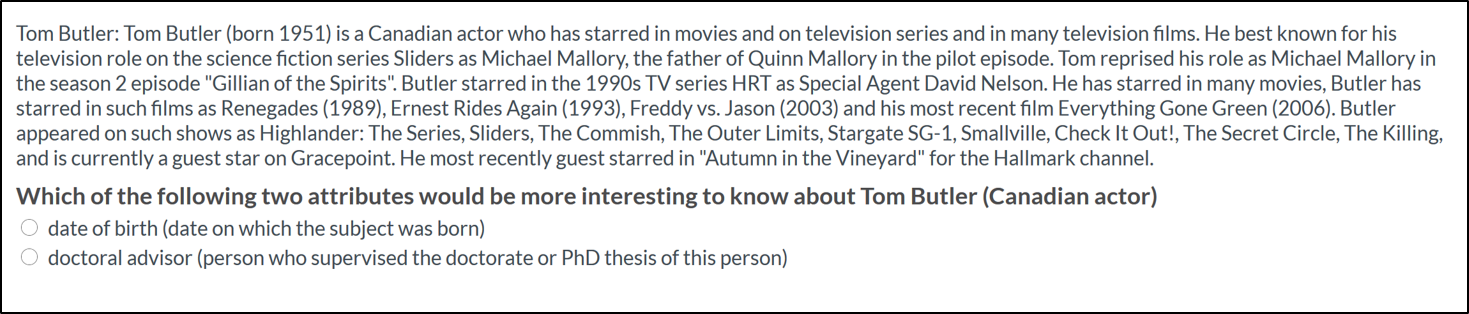}
    \caption{Interface of the crowdsourcing task.}
    \label{fig:crowdinterface}
\end{figure*}

\removeadma{In the following, we describe how we constructed a gold-standard dataset for property interestingness. 
We used the Wikidata knowledge base and the class \emph{human},
which has about 3.4 mio instances.}

\paragraph{Records}
We generated 350 random records consisting of a human and two properties, like (\emph{Trump, doctoral advisor, medical condition}). As sampling humans at random from Wikidata would give mostly unknown persons, we decided to sample humans whose Wikidata pages had been edited within the month of November, 2016. As the pages of famous people tend to get edited more often than the pages of non-famous ones, this gave a better mix of humans of different fame.
As Wikidata items contain a large fraction of properties that are identifiers, with many stemming from national libraries and directories that can only be understood by experts of the respective domain, we did not consider such ID properties\removeadma{ (see discussion in Section~\ref{sec:discussion})}. Of the properties that were not IDs, we considered all that were assigned to humans at least 1000 times, which resulted in 101 as of November 14, 2016.

\paragraph{Annotation}
We used the CrowdFlower platform\footnote{\url{https://www.crowdflower.com}} for obtaining preference judgments. In the annotation task, a short biographical sketch and the two properties were presented to the annotators, and they were asked, knowing about which of the two properties would be more interesting. The core part of the interface is shown in Fig.~\ref{fig:crowdinterface}. Quality was ensured via an entrance test and hidden test questions, based on questions unanimously answered in previous runs. For each record, 10 opinions were collected. At 2 ct.\ per annotation, the platform cost (including fees) to generate the whole dataset was \$100. Six sample records are shown in Table~\ref{tbl:samples}.

\begin{figure}
    \centering
    \includegraphics[width=0.74\columnwidth]{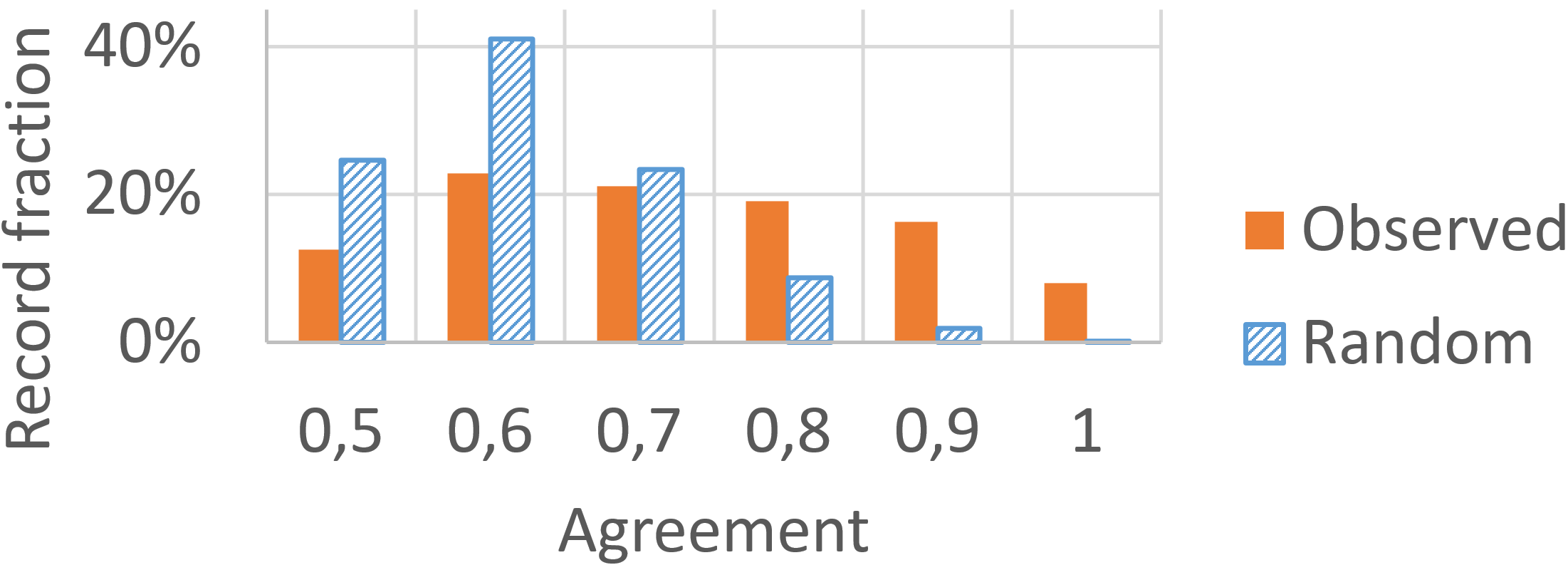}
    \caption{Agreement distribution in our gold dataset.}
    \label{fig:answerdistribution}
\end{figure}

\paragraph{Annotator Agreement}
The agreement of the annotators is shown in Fig.~\ref{fig:answerdistribution} (solid bars). The dashed bars also show the agreement distribution that would be expected if annotators would answer at random. As one can see, annotator agreement is significantly different from random answers, especially evident for the high agreement cases (e.g., if annotators were to answer at random, only 0.2\% instead of 8\% of records would have an agreement of 1). The average agreement is 73\%, and Fleiss' Kappa is 0.40. We also had two authors of this paper annotate a subset of the records, finding that they had 78\% agreement with 16 records where annotators had 80\% agreement, and 100\% agreement with 16 records where annotators had 100\% agreement.
In turn, a manual inspection of low-agreement records showed that many of them correspond to cases where both properties appear unrelated, like \emph{goalsScored} and \emph{militaryRank} for \emph{Pope Francis}, on which agreement is difficult.

\begin{table}[t]
\centering
\resizebox{\textwidth}{!}{%
\begin{tabular}{llllll}
Name                    & Description                 & Property 1         & Property 2                    & Preferred & Agrmt \\ \hline
Albert Johnson & Canadian soccer player & military conflict	& drafted by & Prop. 2 & 1\\
Andrew Collins          & British actor               & field of work      & sister                        & Prop.\ 1   & 0.9        \\
Svetlana Navasardyan & Armenian musician & member of political party & place of detention & \ \ \ \ - & 0.5\\
Filip Stanislaw Dubiski & Polish officer              & residence          & sports discipline & Prop.\ 1   & 1          \\
Dipankar Bhattacharjee  & Indian badminton player     & Roman praenomen    & bowling style                 & Prop.\ 2   & 0.6 \\
David Ball & Electronic music producer & doubles record & military branch & Prop.\ 1 & 0.8\\
Kalim Kashani & 17th century Persian poet & languages spoken/written & sexual orientation & Prop.\ 1 & 1
\end{tabular}}
\caption{Sample records from our gold dataset.}
\label{tbl:samples}
\end{table}

\section{Baselines}
\label{sec:baselines}

\removeadma{In this section we evaluate how well baselines and the state-of-the-art can approximate the annotators' opinions.}

\paragraph{Techniques}

We evaluate four baselines, the first two being simple counts, while the latter two use the state of the art in textual information retrieval and property ranking, respectively:
\begin{enumerate}
\item \emph{Human frequency:} This baseline always chooses the property that is more frequently used for \emph{humans} as winner. \item \emph{Occupation frequency:} A modification of the previous that only looks at people having the same profession, aiming to capture the observation that professions are similar to classes.
\item \emph{Google count:} This method chooses the property with more search results for Google queries concatenating entity name and property as winner. For instance, for \emph{Pope Francis}, \emph{goals scored} and \emph{military rank}, the queries would be ``Pope Francis goals scored'' (470,000 results) and ``Pope Francis military rank'' (3,870,000 results), thus, \emph{military rank} would be chosen as winner. 
\item \emph{Wikidata property suggester:} This baseline uses the suggestions of the Wikidata property suggester, and represents the state of the art in property ranking~\cite{innsbruck:property:ranking}.  The version available online gives natively only few properties, so we modified the code by removing thresholds, in order to obtain further properties. Given an entity-property-property record from our dataset, the winning property according to this method was the one that was ranked higher by the property suggester.
\end{enumerate}

\begin{table}[t]
\centering
\begin{tabular}{l|l|l|l|l}
 & \multicolumn{4}{c}{ppref on records with agreement} \\
Method & \begin{tabular}[c]{@{}l@{}}$\geq$70\%\\ (n=223)\end{tabular} & \begin{tabular}[c]{@{}l@{}}$\geq$80\%\\ (n=150)\end{tabular} & \begin{tabular}[c]{@{}l@{}}$\geq$90\%\\ (n=85)\end{tabular} & \begin{tabular}[c]{@{}l@{}}=100\%\\ (n=28)\end{tabular} \\ \hline
\emph{Random} & 50\% & 50\% & 50 \% & 50\%\\
\emph{Annotators} & 81.8\% & 87.5\% & 93.3\% & 100\%\\ \hline
Human frequency & 57.4\% & 60.6\% & 62.3\% & 50\% \\
Occupation frequency  & 57.4\% & 58.6\% & 61.2\% & 53.6\%  \\
Google count & 58\% & 58.3\% & 61.2\% & 53.6\%\\
Property suggester & 58.7\% & 61.3\% & 62.3\% & 50\%\\ 
\end{tabular}
\caption{Performance of baseline approaches for property ranking.}\label{tbl:baselines}
\end{table}

\paragraph{Evaluation}

For evaluation, we use the \emph{ppref} measure (precision of preference)~\cite{carterette2008here}, henceforth just called precision, which measures the percentage of records where a method proposes the same property as winner as the majority of the annotators does.
\removeadma{}
We present results for records with at least 70\%, 80\%, 90\% and 100\% separately, of which there were 226, 152, 85 and 28, respectively. \removeadma{This allows a comparison of how the methods fare on records where annotators show less or more agreement.} Where trends are similar, we focus in the discussion on the records with at least 80\% agreement. 
If both properties were chosen with equal likelihood, a record would have a probability of 11.2\% to fall into this group, i.e., it is unlikely that for many of the records in this group, the property with more votes has been voted that way only by chance.

\paragraph{Results and Analysis}

As Table~\ref{tbl:baselines} shows, all four baselines perform comparably bad, agreeing only in about 60\% of the answers with the majority of the annotators.
\removeadma{}
For Baselines (1), (2), and (4), we believe their main weakness is that they rely on describing the data that is present (remember \emph{blood type} vs.\ \emph{hair color}, where the former exists 14 times as often as the latter).
While association rules (Baseline 4) are more sophisticated than simple counts (Baselines 1 and 2), apparently, better capturing correlations does not improve predictions on interestingness. It may be noteworthy that association rules were originally developed for discovering patterns in applications where data is complete\removeadma{~\cite{Agrawal:Et:Al-Assoc:Rules-SIGMOD}}, not for making statements about absent data.
\removeadma{}
For Baseline (3), Google count, we trace the low performance to the fact that text search is not able to sufficiently capture the connection between entities and predicates. For instance, for \emph{Pope Francis} and \emph{military rank}, all top ranked results talk about his relation to the Argentinian dictatorship or the Swiss Guard, none about his \emph{own} military rank.



\section{Improving Property Ranking}
\label{sec:advanced}

\removeadma{We next investigate how property ranking can be improved using transfer learning, semantic similarity and ensembles.}

\subsection{Transfer Learning via Property Pivoting}

We have seen that statistical approaches that learn patterns over the whole dataset do not yield a high precision in predicting pairwise interestingness preferences.
We also discussed that it is virtually impossible to obtain enough training data for supervised learning. One approach often taken if supervised learning is not possible is \emph{transfer learning}. Transfer learning refers to the training of models to solve a Problem A, for which enough training data is available, then applying them to a related Problem~B~\cite{transfer-learning}. 
\removeadma{
In our case, Problem B is the pairwise preference between two properties, but what could qualify as Problem A?} An idea adapted from~\cite{bast:relevance:scores:triples} is to predict, which of two properties is asserted for an entity, and which one is not.

Consider the case of \emph{position played on team} (P413) and \emph{religious order} (P611). There are 225,821 humans in Wikidata that have the former property but not the latter, and 7,976 humans that have the latter but not the former. Are there chances to accurately decide, for a person picked from either of these sets, to which of the two sets it belongs? We call this the \emph{property pivoting problem}.

\paragraph{Deciding Property Pivoting via Regression}
In the following, we use a logistic regression classifier trained on bags of words taken from person descriptions from Wikipedia to decide the property pivoting problem.

Wikipedia articles provide a messier, but considerably larger source of information about a person than Wikidata. By considering Wikipedia articles as bags of words, we can use the number of occurrences of each word as feature on which to train a classifier. In particular, for each pair of properties, we trained a logistic regression classifier on up to 20,000 entities, if available: 10,000 entities that had the first property but not the second one, and 10,000 entities that had the second but not the first property.

\removeadma{
For the input to the regression, we did a basic preprocessing, in particular punctuation and stop word removal, case conversion and stemming. It is then both possible to give the true word count as input (unweighted), or to use a weighting scheme such as TF-IDF to give more importance to words that occur less frequent. We experimented with both variants, for TF-IDF, as usual, removing the top and bottom 20\% of words, as these mostly correspond to proper names that occur too seldom to carry any value, or to very unspecific general words.
}

As an example of what these classifiers learned, the box below shows the most distinctive weights learned by the TF-IDF-based regression classifier that was trained for \emph{position played on team} versus \emph{religious order}:

\smallskip

\begin{center}
\begin{tabular}{|llll|}
\hline
-3.09: footballer & -2.39: season\ \ \ \ \ \ & 2.67: jesuit & 1.88: work\\
-2.85: football & -2.21: career & 2.43: catholic & 1.85: life\\
-2.75: played & -2.16: league &  2.41: died & 1.77: order \\
-2.72: team & -2.08: cup & 2.04: priest & 1.64: church\\
-2.67: player & -2.02: club & 1.99: works & 1.50: death \\
\hline
\end{tabular}
\end{center}

\smallskip
\noindent

Negative weights hereby indicate that the occurrences of the word appear frequently in articles of entities with \emph{position played on team} being present instead of \emph{religious order}, while positive weights indicate the opposite. For instance, \emph{season} (weight -2.39) is a word in the former category, while \emph{priest} (weight (2.04) belongs to the latter. Interestingly, also some rather general words like \emph{works} and \emph{died} appear in the latter category. We conjecture that soccer players are mostly recent figures still alive, while monks are more frequently from the past, thus, \emph{died} is more relevant for them, and that monks are known for more diverse activities than sports players, thus the term \emph{work}.

\paragraph{Property Pivoting Quality}

Evaluated each over 200 entities that had either the one or the other property,
the regression classifiers achieved a respectable average precision of 94.8\%. For \emph{position played on team} versus \emph{religious order}, for instance, the precision is even 100\%, while difficult cases are for instance \emph{field of work} versus \emph{member of} (84\% precision), or \emph{child} versus \emph{sister} (72\% precision).

\paragraph{Transfer Learning}
Our hope was that characteristics that describe whether a person has one property, but not another also relate to how interesting the one property is over the other. That is, for people that have a \emph{position played on team} but no \emph{religious order}, maybe knowing about the former is indeed more interesting than the latter. The results of transferring our regression classifiers are shown in Table~\ref{tbl:advanced} (third and fourth row). As we can see, the precision on records with at least 80\% annotator agreement is 69.3\% and 72\%, depending on whether TF-IDF is used or not. These precisions are remarkably better than those of the baselines, though still leaving a considerable gap to the 95\% precision on the pivoting task, indicating that property pivoting and property interestingness are only moderately related problems. For instance, although all annotators agree that for the soccer player \emph{Albert Johnson}, \emph{drafted by} is more interesting than \emph{military conflict} (Table~\ref{tbl:samples} first row), property pivoting still chooses the latter property, presumably, because \emph{drafted by} is used in Wikidata only in very specific contexts of baseball and ice hockey, not for soccer. Similarly, for Kalim Kashani, a 17th century Persian poet, property pivoting chooses \emph{sexual orientation} over \emph{languages spoken or written}, although crowd agreement is 100\% on \emph{languages spoken or written} (Table~\ref{tbl:samples} last row). Presumably the reason is that many people that are not poets have information about languages as well, while sexual orientation is especially frequently asserted for artists.

\subsection{Semantic Approaches}

As exemplified above, while regression is very well able to decide the property pivoting problem, existence of a property is still only a moderate indicator for interestingness. In particular, there are several cases where it is intuitive that one property does not apply to a person at all, while another is relevant, but regression is not able to discover that.
\removeadma{}
In the following we thus propose to use semantic similarity 
as alternative proxy for interestingness. We conjecture that if a property bears some semantic similarity to an entity, like
%
\emph{goals scored} for \emph{Ronaldo}, where humans can easily see an association of Ronaldo scoring goals, then it is more likely that that property is also interesting.

Concretely, we propose to look at the semantic similarity of the textual descriptions of entities and properties.
For entities, we use their English Wikipedia articles as text sources, while for properties we use their textual label and description on Wikidata. 
\removeadma{For the entities in our dataset, this results in textual descriptions typically consisting of at least 5 sentences, while properties are described on Wikipedia typically in a single phrase, like \emph{``goals/points scored in a match or an event used as qualifier to the participant}'' for the property \emph{goals scored.}\footnote{\url{https://www.wikidata.org/wiki/Property:P1351}}
}
To compute semantic similarity, we rely on standard latent topic models, in particular Latent Semantic Indexing (LSI)~\cite{lsi} and Latent Dirichlet Allocation (LDA)~\cite{lda}. We proceed in three steps:
\begin{enumerate}
	\item In the first step, we train topic models on Wikipedia.
	\item In the second step, we represent each entity and each property as distribution over the learned topics.
	\item In the third step, for each (entity, property, property) record in our gold dataset, we compute the similarity between the topics of the entity and the two properties. We then assert that the more similar property is the more interesting one.
\end{enumerate}
We further detail each step below.

\paragraph{Learning Topic Models}

In the first step, we used LSI and LDA to learn 400 and 100\removeadma{,\footnote{We also tested LDA at 400 topics, but observed no improvement} respectively,} distinct topics over the English Wikipedia text corpus (12.5 GB). \removeadma{A topic by itself is a distribution over words, i.e., it consists of a set of weights for words in the lexicon, describing how frequently each word occurs in that topic.}
\removeadma{}
We used the \textit{gensim} Python library, with which training could be done on a standard laptop within hours.
LDA has a parameter $\alpha$, which is a prior asserting whether texts are preferentially assigned to more or to fewer topics. This parameter is important for our application as assignment to fewer topics leads to sparser vectors. We report the results for the default value (1/\#topics=0.01). We also tested a higher value, 0.05, and a setting called auto-optimization, where $\alpha$ is dynamically set for each topic, but both performed worse. Notably, we found in this setting that too high values of $\alpha$ lead to properties being assigned all to the same topics, resulting in entity-property distances that were indistinguishable.
\removeadma{}
In Table~\ref{tbl:topics} we show some of the most frequent words in four topics as learned by LDA.
\removeadma{We found that most learned topics appeared coherent to a human observer. There were, though, a few weird mergers of topics, for instance a topic containing mostly words about Myanmar and Nottingham. Also, a few topics were not accessible to us, and appeared to be collections of residual words not fitting in the other topics.}

\begin{table}[t]
\centering
\includegraphics[width=\textwidth]{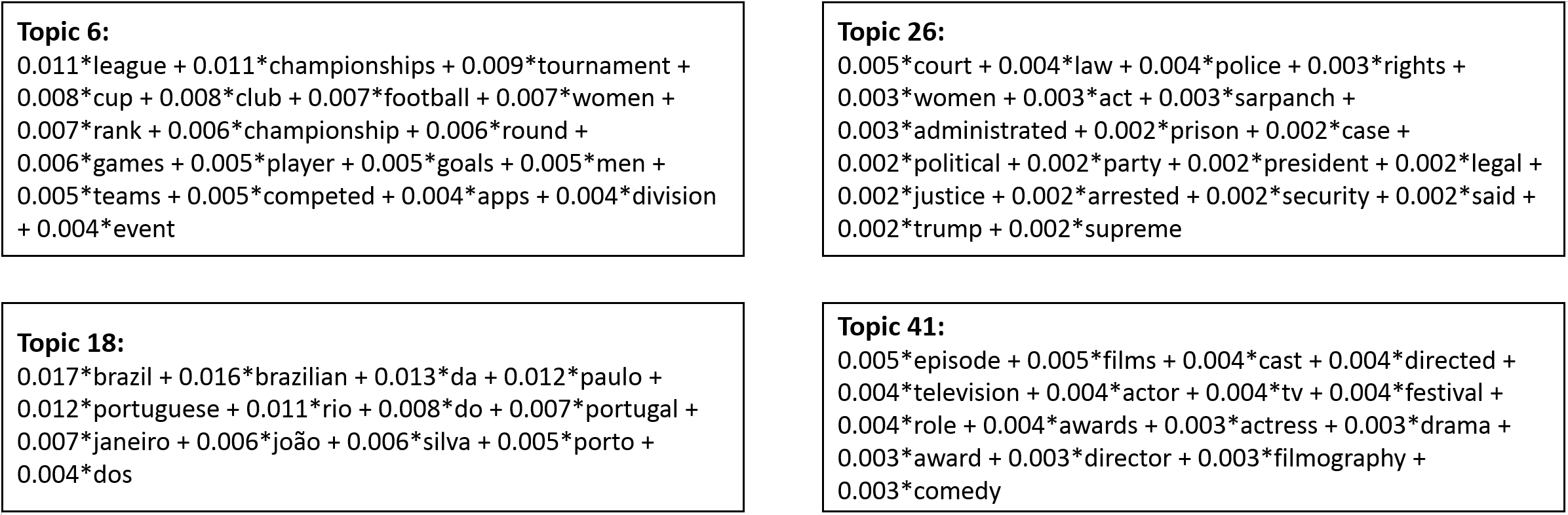}
\caption{Sample topics as learned by LDA over Wikipedia}
\label{tbl:topics}
\end{table}

\paragraph{Describing Entities and Properties using Topic Models}

In the next step we used the learned topic models to describe entities and properties. To that end, given a set of words, LSI and LDA are able to compute a distribution of topics that is the most likely one to generate the given set of words.

For the soccer player \emph{Ronaldo}, for instance, LDA states that his Wikipedia article can be described using 52\% of Topic 6, 12\% of Topic 18, 7\% of Topic 41, 6\% of Topic 26 (all shown in Table~\ref{tbl:topics}), and low fractions of a few others.
This distribution appears sensible, as he is most importantly known for playing soccer in leagues and tournaments (Topic 6), comes from Portugal (Topic 18, note that LDA has merged Brazil and Portugal into one topic), is featured frequently in the media (Topic 41) and 
has been under legal investigation (Topic 26).
Similarly, the article of the former US president \emph{Barack Obama} can be generated by combining 48\% of Topic 26 (law and politics), 18\% of a topic concerned with parties and elections, 12\% of a topic concerned with business and industry, and various others.

In the same way, also properties can be described as combinations of topics. The property \emph{height}, for instance, is composed of Topic 14 (geometry), 36 (abstraction), 84 (biology) and an unclear topic 88, whereas \emph{member of sports team} is composed entirely of Topic 6.

\paragraph{Computing Similarities}

For computing similarities between entities and properties, we use cosine similarity between vectors, a standard approach in vector-space-modelling and word embedding\removeadma{~\cite{similarity-topic-models}}.
The idea is to interpret topic distributions  as vectors in a high-dimensional space, then compute the distance using the cosine of the angle between these vectors. In the case of \emph{Ronaldo} and \emph{goals scored} versus \emph{military rank}, for instance, we find a cosine similarity of 0.987 versus 0.611. Thus, \emph{goals scored} is the semantically more similar property, and hence we propose this to be the more interesting one.
\removeadma{Similarly, for \emph{Kalim Khasani}, we find the cosine wrt.\ \emph{languages spoken/written} to be 0.108, and wrt.\ \emph{sexual orientation} to be 0.004, thus \emph{languages spoken/written} would be considered more interesting.}

\paragraph{Analysis}

The performance of semantic similarity as a proxy for interestingness is shown in Table~\ref{tbl:advanced}. As we can see, LDA does not outperform the baselines at 60\% precision for records with at least 80\% annotator agreement. In contrast, LSI performs considerably better at 65.3\% precision, though still not achieving the performance of regression (72\%). 

We find that semantic similarity is better able to capture when one property does not make sense at all, as evidenced by the increase (LDA) or smaller drop (LSI) towards the records with 100\% human agreement than regression. For instance, from the records with at least 80\% agreement, the precision of LDA increases by 11.4\%, the precision of LSI decreases by 1.1\%, but the precision of regression drops by 6.1/7.5\%.


But there are also various spectacular failures: 
For \emph{Gabriel Kicsid}, a handball player, for instance, both LSI and LDA believe that \emph{religious order} is more similar than \emph{follows}, even though all annotators agree that the second is more interesting. A possible reason is that the description of \emph{follows} is quite abstract and hard to match (``\emph{immediately prior item in some series of which the subject is part}''\footnote{\url{https://www.wikidata.org/wiki/Property:P155}}), even though human annotators correctly understand the usage, which is about succession for instance on a position in a team.
Similarly, for the Polish officer \emph{Filip Stanislaw Dubitzki} (Table~\ref{tbl:samples}, 4th row), unlike all annotators, LSI
believes that \emph{sports discipline} is more interesting than \emph{residence}.

\removeadma{
It appears that a challenge for both methods lies in the shortness of the descriptions, both of properties and entities, which makes the correct identification of topics challenging. The property \emph{employer}, for instance, is assigned to only two topics by LDA, the property \emph{goals scored} to even just one. Similarly, entities with short descriptions are assigned to only 4-8 topics. Then, when for a given entity topic vector, the topic vectors of both properties to be ranked are nearly orthogonal, the ranking becomes unreliable. And even though the topic distribution can be influenced using the parameter $\alpha$ for LDA, it appears that favoring broader distribution (i.e., more topics per entity/property) does not help, as the text sources do not allow to identify enough distinguishable topics.
}


\subsection{Ensembles}

\begin{table}[t]
\centering
\begin{tabular}{l|llllllll}
                      & \rot{Human freq.} & \rot{Occupation freq.} & \rot{Google count} & \rot{Property Suggester} & \rot{Regression (plain)} & \rot{Regression (TF-IDF)} & \rot{LDA} & \rot{LSI} \\
                      \hline
Human frequency & - & 0.37  & 0.11    & {0.99}    & 0.29  & 0.33            & -0.07  & 0.07        \\
Occupation frequency     & 0.37         &         -          & -0.02                 & 0.36              & 0.33             & 0.25            & 0.08        & 0.00        \\
Google count & 0.13         & -0.02             &           -            & 0.11              & 0.01             & 0.03            & -0.10       & -0.01       \\
Property Suggester     & {0.99}        & 0.36              & 0.11                  &         -          & 0.28             & 0.33            & -0.07       & 0.05        \\
Regression (unweighted)     & 0.26         & 0.33              & 0.01                  & 0.28              &         -         & {0.87}            & 0.14        & 0.29        \\
Regression (TF-IDF)      & 0.32         & 0.25              & 0.03                  & 0.33              & {0.87}             &          -       & 0.09        & 0.26        \\
LDA         & -0.09        & 0.08              & -0.10                 & -0.07             & 0.14             & 0.09            &        -     & 0.16        \\
LSI          & 0.04         & 0.00              & -0.01                 & 0.05              & 0.29             & 0.26            & 0.16        &  -          
\end{tabular}
\caption{Pearson correlation coefficients on records with at least 80\% agreement}
\label{tbl:correlations}
\end{table}

Considering the methods discussed so far along with their strong and weak points, the question arises whether it is possible to combine them and achieve a better performance. By Condorcet's jury theorem~\cite{jury-theorem}, it is beneficial to combine weak predictors whenever these show a sufficient level of statistical independence, and have each more than 50\% accuracy. To see whether our methods exhibit sufficient statistical independence, we computed pairwise Pearson correlation coefficients, which are shown in Table~\ref{tbl:correlations}. 

There are two surprising insights. The first is the near-perfect correlation between human frequency and Property Suggester (0.99). In other words, even though the Property Suggester uses sophisticated ways for computing association rules and combining predictions, it does essentially nothing different from simply counting how often a property occurs. 
The second surprising insight is that apart from human frequency/property suggester and the two variants of regression, the pairwise correlation between the methods is very low. This even holds for similar techniques such as human frequency compared with occupation frequency (correlation 0.37) or LSI compared with LDA (correlation 0.16). Two methods, Google count and LDA, exhibit almost no correlation to any other method. 


The results are in itself remarkable, and suggest that ensembles predictors\footnote{Not to be mixed with \emph{ensemble learning}, a machine learning approach where consecutive instances of the same classifier are trained especially on records that previous instances predicted wrongly. Ensemble learning requires a sufficient amount of labeled training data, which is not available in our case.}  can give better performances than the individual methods. We tested to take the majority vote from various permutations of three and five of our presented methods. It turned out that the best performing ensemble was not only a combination of three advanced methods, but that even adding some of the baselines improved precision. In particular, the best performing combination used five methods that showed the biggest pairwise difference in Pearson correlation, namely the Google count, LSI, LDA, Occupation frequency, and regression (TF-IDF). This ensemble performed 2\% and 4.6\% better than the best single method, regression (TF-IDF) on records with at least 80\% and 90\% agreement, respectively. 


\subsection{Analysis}

Results on the performance of the baselines are shown in Table~\ref{tbl:baselines}, while the performance of the advanced methods is shown in Table~\ref{tbl:advanced}. In both cases, we can see random agreement (50\%) as a lower bound that any method should outperform, and annotator agreement as an upper bound. As one can see, regression trained on property pivoting as the best single advanced method beats the baselines by a margin of 10\% on records with at least 80\% agreement,  with LSI performing 5\% worse, and LDA being on par with the baselines. The best ensembles then add a further 2\% precision on top of the regression. We draw the following conclusions:
\begin{enumerate}
	\item \emph{The state of the art methods alone are inadequate for property ranking.} The state of the art in property suggestion (Property Suggester) and document retrieval (Google count) achieved only 58.3 and 61.3\% precision on records where annotators had 87.5\% agreement, not significantly different from an approach that simply counts how often a property appears (60.6\% precision).
	\item \emph{Regression is well-suited for property pivoting, but has still limitations, as it is data-driven.} Regression based on bags-of-words from Wikipedia articles achieves an accuracy of 94.8\% for property pivoting on records with at least 80\% agreement, i.e., deciding whether an entity has one property but not another. And while it is an expensive method, requiring to train $O(n^2)$ many classifiers for $n$ properties, it is worth its price also for property ranking, as it outperforms the baselines by more than 10\%.
\removeadma{}	
	It has still limitations as it is data-driven, though, i.e., it predicts interestingness of properties based on their presence, which in some cases is not a good indicator.
	\item \emph{Semantic approaches are great at discovering applicability of properties, but struggle with short property descriptions.} Semantic similarity based on latent topic modeling turned out to be better able to capture cases where certain properties did not at all make sense. Nevertheless, 
    there are problems with description shortness.
    \removeadma{we found that there were problems with the shortness of property descriptions, and that predicting interestingness purely based on similarity misses an important aspect of inherent importance.}
	\item \emph{Ensembles work best.} 
    Taking the majority vote among methods based on counting (Occupation frequency, Google count), correlation (regression) and semantic similarity (LSI, LDA) 
    approximated human judgment best.
\end{enumerate}

\begin{table}[t]
\centering
\resizebox{0.6\textwidth}{!}{%
\begin{tabular}{l|l|l|l|l}
 & \multicolumn{4}{c}{ppref on records with agreement} \\
Method & \begin{tabular}[c]{@{}l@{}}$\geq$70\%\\ (n=223)\end{tabular} & \begin{tabular}[c]{@{}l@{}}$\geq$80\%\\ (n=150)\end{tabular} & \begin{tabular}[c]{@{}l@{}}$\geq$90\%\\ (n=85)\end{tabular} & \begin{tabular}[c]{@{}l@{}}=100\%\\ (n=28)\end{tabular} \\ \hline
\emph{Random} & 50\% & 50\% & 50\% & 50\%\\
\emph{Annotators} & 81.8\% & 87.5\% & 93.3\% & 100\%\\ \hline
\begin{tabular}[c]{@{}l@{}}Regression (unweighted) \end{tabular} & 67.7\% & 69.3\% & 70.4\% & 64.3\%\\
\begin{tabular}[c]{@{}l@{}}Regression (TF-IDF) \end{tabular} & 70.4\% & 72\% & 71.8\% & 64.3\% \\ 
LDA & 57\% & 60\% & 61.2\% & 71.4\% \\ 
LSI & 59\% & 65.3\% & 67\% & 64.3\%\\








\begin{tabular}[c]{@{}l@{}}Ensemble  of Google count,LSI,\\ LDA, Occupation frequency,\\ regression (TF-IDF)
\end{tabular} & 69.1\% & 74\% & 76.4\% & 67.9\% \\
\end{tabular}}
\caption{Performance of advanced methods for property ranking.}
\label{tbl:advanced}
\end{table}

\removeadma{

\section{Discussion}
\label{sec:discussion}

In the following we discuss some practical aspects of our experiment setup.




\paragraph{Interestingness and Applicability}

An issue that kept our attention was the phrasing of the question posed to the crowd workers, \emph{``Which of the two properties would be more interesting to know about.''}

A concern was that certain properties would be interesting if they existed, and that the question phrasing could be understood as a question for interestingness if the property existed. For instance, while football players are rarely members of a monastic order, if one of them is, then \emph{religious order} would be interesting. Similarly, most musicians have not been detained anywhere, but if a musician has been detained then \emph{place of detention} could be interesting.

While it is tempting to thus ask separate questions \emph{``1. How likely is it that a value for this property exists,''} and \emph{``2. How interesting would it be to know this value,''} this does not solve the problem, as in some cases also the absence of values is interesting, for instance, that Angela Merkel has no children, or that the scientist Michael Faraday had no university education.

Inspection of the dataset shows that crowd workers got the issue generally right, and answered how interesting it would be to learn about a property, not presuming that a value actually exists (see e.g., Record 1 in Table~\ref{tbl:samples}, where \emph{military conflict} did not receive any votes).

\paragraph{Property Semantics}

A challenge for data-driven methods was that some properties were only used in narrow contexts, for instance, \emph{drafted by} is only used in baseball and ice hockey, or \emph{singles record} is only used for scores in sports tournaments. Human annotators, in contrast, found that \emph{drafted by} made also sense for soccer players, or that \emph{singles record} could also refer to the number of singles produced/sold by an electronic music producer. As all semantics in Wikidata is derived by consens, and the data model has virtually no hard constraints, the monitoring of descriptions, intended use and actual use of properties will pose a continuous challenge.

\paragraph{IDs}

Wikidata articles often contain also a considerable set of ID properties, ranging from IDs stemming from the library domain (VIAF ID, GND ID, SUDOC ID) to web directory and service IDs (Facebook and Twitter username, Rotten Tomatoes ID, \ldots), that typically make up 10\% to 40\% of all properties of persons.

We filtered out ID properties from our dataset, because ranking especially the bibliographic IDs can only be reasonably done by library domain experts. Nevertheless, we are aware that one should not neglect their importance, especially the Virtual International Authority File (VIAF) is a major international effort to uniquely identify persons across a large set of national identifiers.

Similarly, some web specific IDs are prime pieces of information (imagine the page of Donald Trump missing information about his Twitter usernames), and their importance should not be underestimated.

}

\section{Conclusion}

We have introduced the problem of property ranking, shown the limitations of the state-of-the-art, and developed approaches that combine classical frequency-based approaches, transfer learning and semantic similarity. Our methods outperform the state of the art by over 10\% precision, though still being inferior to human agreement by 11.5\%. We hope that the dataset developed in this paper can stimulate research that can further approach human agreement.

We see two interesting avenues to extend this work: One is to improve the methods presented in this paper, for instance, by using other learning algorithms for the property pivoting problem, or by extending the short descriptions from which the semantic methods currently learn. The other is to find completely new approaches to the problem, which, even if they do not individually outperform the existing methods, might add information to ensembles. The challenge would be here to find related problems that can be used for transfer learning.







\bibliographystyle{plain}

{
\bibliography{recommendations}

\begin{thebibliography}{10}

\bibitem{DBLP:journals/dbsk/AbedjanN13}
Ziawasch Abedjan and Felix Naumann.
\newblock Improving {RDF} data through association rule mining.
\newblock {\em Datenbank-Spektrum}, 13(2):111--120, 2013.

\bibitem{Agrawal:Et:Al-Assoc:Rules-SIGMOD}
Rakesh Agrawal, Tomasz Imielinski, and Arun~N. Swami.
\newblock Mining association rules between sets of items in large databases.
\newblock In {\em {SIGMOD} 1993.}, pages 207--216, 1993.

\bibitem{recoin}
Albin Ahmeti, Simon Razniewski, and Axel Polleres.
\newblock Assessing the completeness of entities in knowledge bases.
\newblock In {\em ESWC P\&D}, 2017.

\bibitem{reverse-engineer-properties}
Ahmad Assaf, Ghislain~A Atemezing, Rapha{\"e}l Troncy, and Elena Cabrio.
\newblock What are the important properties of an entity?
\newblock In {\em ESWC}, pages 190--194, 2014.

\bibitem{dbpedia}
S{\"o}ren Auer, Christian Bizer, Georgi Kobilarov, Jens Lehmann, Richard
  Cyganiak, and Zachary Ives.
\newblock {\em {DBpedia}: A nucleus for a web of open data}.
\newblock 2007.

\bibitem{bast:relevance:scores:triples}
Hannah Bast, Bj\"{o}rn Buchhold, and Elmar Haussmann.
\newblock Relevance scores for triples from type-like relations.
\newblock In {\em SIGIR}, pages 243--252, New York, NY, USA, 2015.

\bibitem{lda}
David~M Blei, Andrew~Y Ng, and Michael~I Jordan.
\newblock Latent {D}irichlet allocation.
\newblock {\em Journal of machine Learning research}, 3(Jan):993--1022, 2003.

\bibitem{learningtorank}
Zhe Cao, Tao Qin, Tie-Yan Liu, Ming-Feng Tsai, and Hang Li.
\newblock Learning to rank: From pairwise approach to listwise approach.
\newblock In {\em ICML}, pages 129--136, 2007.

\bibitem{carterette2008here}
Ben Carterette, Paul~N Bennett, David~Maxwell Chickering, and Susan~T Dumais.
\newblock Here or there: Preference judgments for relevance.
\newblock In {\em ECIR}, pages 16--27, 2008.

\bibitem{jury-theorem}
Marquis~de Condorcet.
\newblock Essai sur l'application de l'analyse \`a la probabilit\'e des
  d\'ecisions rendues \`a la pluralit\'e des voix.
\newblock {\em Paris: Imprimerie Royale}, 1785.

\bibitem{lsi}
Scott Deerwester.
\newblock Improving information retrieval with latent semantic indexing.
\newblock 1988.

\bibitem{dessi:atzori:ranking:properties}
Andrea Dessi and Maurizio Atzori.
\newblock A machine-learning approach to ranking {RDF} properties.
\newblock {\em FGCS}, 54:366--377, 2016.

\bibitem{unusual-suspects}
Nausheen Fatma, Manoj Chinnakotla, and Manish Shrivastava.
\newblock The unusual suspects: Deep learning based mining of interesting
  entity trivia from knowledge graphs.
\newblock In {\em AAAI 2017}, 2017.

\bibitem{gamon-text}
Michael Gamon, Arjun Mukherjee, and Patrick Pantel.
\newblock Predicting interesting things in text.
\newblock In {\em COLING}, pages 1477--1488, 2014.

\bibitem{gassler2014guided}
Wolfgang Gassler, Eva Zangerle, and G{\"u}nther Specht.
\newblock Guided curation of semistructured data in collaboratively-built
  knowledge bases.
\newblock {\em FGCS}, 31, 2014.

\bibitem{wsdm-cup}
Stefan Heindorf, Martin Potthast, Hannah Bast, Bj{\"{o}}rn Buchhold, and Elmar
  Haussmann.
\newblock {WSDM} cup 2017: Vandalism detection and triple scoring.
\newblock 2017.

\bibitem{pointwise-not-good}
Nicolas Jones, Armelle Brun, and Anne Boyer.
\newblock Comparisons instead of ratings: Towards more stable preferences.
\newblock In {\em WI-IAT}, pages 451--456, 2011.

\bibitem{pointwise-not-good-2}
Saikishore Kalloori, Francesco Ricci, and Marko Tkalcic.
\newblock Pairwise preferences based matrix factorization and nearest neighbor
  recommendation techniques.
\newblock 2016.

\bibitem{schema-induction-2-probabilistic}
Kenza Kellou-Menouer and Zoubida Kedad.
\newblock Schema discovery in {RDF} data sources.
\newblock In {\em International Conference on Conceptual Modeling}, pages
  481--495, 2015.

\bibitem{deswebe-2014-hpi}
Philipp Langer, Patrick Schulze, Stefan George, Matthias Kohnen, Tobias Metzke,
  Ziawasch Abedjan, and Gjergji Kasneci.
\newblock Assigning global relevance scores to dbpedia facts.
\newblock In {\em ICDE workshops}, pages 248--253, 2014.

\bibitem{lee2013attribute}
Taesung Lee, Zhongyuan Wang, Haixun Wang, and Seung-won Hwang.
\newblock Attribute extraction and scoring: A probabilistic approach.
\newblock In {\em ICDE}, pages 194--205, 2013.

\bibitem{l2r2}
Hang Li.
\newblock A short introduction to learning to rank.
\newblock In {\em IEICE Transactions}.

\bibitem{ibminer}
Hamid Mousavi, Shi Gao, and Carlo Zaniolo.
\newblock {IBminer}: A text mining tool for constructing and populating infobox
  databases and knowledge bases.
\newblock {\em VLDB}, 2013.

\bibitem{transfer-learning}
Sinno~Jialin Pan and Qiang Yang.
\newblock A survey on transfer learning.
\newblock {\em TKDE}, 2010.

\bibitem{prakash2015did}
Abhay Prakash, Manoj~Kumar Chinnakotla, Dhaval Patel, and Puneet Garg.
\newblock Did you know?-mining interesting trivia for entities from wikipedia.
\newblock 2015.

\bibitem{vision}
Simon Razniewski, Fabian~M Suchanek, and Werner Nutt.
\newblock But what do we actually know.
\newblock {\em AKBC}, pages 40--44, 2016.

\bibitem{similarity-topic-models}
Mark Steyvers and Tom Griffiths.
\newblock Probabilistic topic models.
\newblock {\em Handbook of latent semantic analysis}, 427(7):424--440, 2007.

\bibitem{yago}
Fabian~M Suchanek, Gjergji Kasneci, and Gerhard Weikum.
\newblock {YAGO}: a core of semantic knowledge.
\newblock In {\em WWW}, pages 697--706, 2007.

\bibitem{schema-induction}
Johanna V{\"o}lker and Mathias Niepert.
\newblock {\em Statistical Schema Induction}.
\newblock 2011.

\bibitem{wikidata}
Denny Vrande{\v{c}}i{\'c} and Markus Kr{\"o}tzsch.
\newblock Wikidata: a free collaborative knowledge base.
\newblock {\em CACM}, 57(10):78--85, 2014.

\bibitem{innsbruck:property:ranking}
Eva Zangerle, Wolfgang Gassler, Martin Pichl, Stefan Steinhauser, and Günther
  Specht.
\newblock An empirical evaluation of property recommender systems for
  {Wikidata} and collaborative knowledge bases.
\newblock In {\em Opensym}, 2016.

\end{thebibliography}
}

\appendix

\removeadma{
\section*{Appendix}

\section{Inapplicability of Pointwise Interestingness Scores}
\label{appendix-frameproblem}

A challenge with the Wikidata knowledge base is that its structure is constantly evolving, in particular, that the set of properties is growing over time. A conjecture was that this may make the interpretation of pointwise interestingness scores difficult over time, as interestingness would in part depend on ``what else'' one could express as information for entities.

To verify whether pointwise annotations of property interestingness are stable, 
 we conducted the following experiment:

We chose two soccer players, one famous and one little known one (\emph{Cristiano Ronaldo} and \emph{Akaki Gogia}), and fixed 3 properties for benchmarking, which were \emph{sport, website} and \emph{spouse}. Furthermore, we selected 7 properties of high importance (\emph{Country of origin, participant of, award received, date of birth, member of sports team, position played on team / speciality, country of citizenship}) and 7 properties of low importance (\emph{academic degree, image of grave, educated at, brother, hair color, military rank, religion}) to be used as varying contexts.

We then generated two setups: In Setup 1, crowd workers were given a batch of tasks showning first the 7 important, then the 3 benchmark properties, and asked for each of them for a 0/1 rating of importance. In Setup 2, crowd workers were shown the 7 properties of low importance first, then the 3 benchmark properties.

The setups were given to 20 crowdworkers each. For each of the 6 player-property pairs (Ronaldo/Gogia and sport/website/spouse), we then measured the difference in interestingness as stated by the crowd. The results are shown in Table~\ref{fig:framedependency}. In each of the 6 cases, the properties achieved a higher interestingness when presented with the 7 unimportant ones first, with an average score difference of 0.19 per pair. For instance, for Ronaldo, the property spouse received an interestingness score of 0.35 in Setup 1, and 0.50 in Setup 2. In total, out of the 120 votes given in each setup, in Setup 1, 51 were given for interesting, while in Setup 2, 74 were given for interesting. This leads us to the conclusion that pointwise interestingness scores are highly dependent on the context in which they are presented, and that without specific setups that ensure stability of context, pointwise scores collected today cannot be interpreted easily interpreted tomorrow.

\begin{table}
\centering
\begin{tabular}{ll|lll}
        &         & Setup 1 & Setup 2 & Difference \\ \hline
        & sport   & 0.65    & 0.80    & +0.15       \\
Ronaldo & website & 0.45    & 0.75    & +0.30       \\
        & spouse  & 0.35    & 0.50    & +0.15       \\
        & sport   & 0.60    & 0.65    & +0.05       \\
Akaki   & website & 0.40    & 0.60    & +0.20       \\
        & spouse  & 0.15    & 0.40    & +0.25       \\ \hline
        & Average & 0.43    & 0.62    & +0.19      
\end{tabular}
\caption{Interestingness scores in varying contexts.}
\label{fig:framedependency}
\end{table}

}





\end{document}